\documentclass[aps, prl,twocolumn,10pt,showpacs, floatfix]{revtex4-1}
\usepackage{amsmath}
\usepackage{graphicx}
\usepackage{graphics}
\usepackage{units}
\usepackage{color}
\usepackage{xcolor}
\usepackage[utf8]{inputenc}
\usepackage{textcomp}

\def \CIGSe {Cu(In,Ga)Se$_{\sss 2}$}
\def \CISe {CuInSe$_{\sss 2}$}
\def \CGSe {CuGaSe$_{\sss 2}$}
\def \S3 {$\Sigma 3$}

\def \sss{\scriptscriptstyle\rm}

\begin{document}

\title{Behavior of $\mathbf{\Sigma 3}$ grain boundaries in \CISe\ and \CGSe\ photovoltaic absorbers 
as revealed by first-principles hybrid functional calculations }%

\author{Hossein Mirhosseini}
\email{mirhosse@cpfs.mpg.de}
\affiliation{Max-Planck-Institut f\"ur Chemische Physik fester Stoffe, 01187 Dresden, Germany }

\author{Janos Kiss}
\affiliation{Max-Planck-Institut f\"ur Chemische Physik fester Stoffe, 01187 Dresden, Germany }

\author{Claudia Felser}
\affiliation{Max-Planck-Institut f\"ur Chemische Physik fester Stoffe, 01187 Dresden, Germany }

\begin{abstract}
The inconclusive results of the previous first-principles studies on the \S3 \ 
grain boundaries (GBs) in \CISe\ reveal the importance of employing a method 
that can correctly describe the electronic structure of this solar-cell 
material. We have employed hybrid functional calculations to study the \S3 (112) 
and \S3 (114) GBs in \CISe\ and \CGSe. The electronic structure changes 
introduced by the formation of GBs are threefold: creation of gap states, shift 
in band edges, and alteration of bandgap sizes.  Gap states commonly behave as 
recombination centers, but the band alignment and the change in the bandgap size 
induced by GBs mitigate the destructive effect of these states in \CISe. That 
means, \S3 \ GBs are not detrimental for the carrier transport in devices based 
on \CISe. Conversely, these GBs are destructive for the carrier transport in 
\CGSe. The different behaviors of the \S3 \ GBs in CISe and CGSe might be considered 
by experimentalists to optimize the device fabrication to achieve high-performance 
solar cells. 
\end{abstract}

% 61.72.Mm 	Grain and twin boundaries
% 71.15.Mb 	Density functional theory, local density approximation, gradient and other corrections
% 61.72.Bb 	Theories and models of crystal defects

% 73.20.-r 	Electron states at surfaces and interfaces
%\keywords{}

\maketitle
\section{Introduction}
Thin-film solar cells based on \CIGSe\ (CIGSe) are fabricated and deployed 
worldwide on an industrial scale due to their outstanding price/performance 
ratio~\cite{chirila_NM_2013, jackson_PSSRRL_2015}. In contrast to silicon-based 
solar cells~\cite{seager_PRL_1979}, the efficiency of CIGSe cells with 
polycrystalline light absorber exceeds the efficiency of their monocrystalline 
counterpart~\cite{rau_APA_2009}. This is remarkable, because typically the 
performance of optoelectronic devices is considered to be worse for 
polycrystalline semiconductors due to the presence of grain boundaries (GBs). 
GBs are expected to create deep gap levels that act as recombination centers and 
are commonly regarded detrimental for the solar-cell performance. Despite the 
extensive studies carried out in the past decades to investigate the effects 
of GBs on the efficiency of CIGSe-based solar cells, their role remains a topic 
of intense debate. 

In the literature, different atomic structures for symmetric GBs in \CISe\ and 
\CGSe\ have been reported. Abou-Ras~\emph{et al.}~\cite{abou-ras_PRL_2012} 
identified and studied experimentally two types of \S3 (112) GBs in \CISe: 
Se-Se-terminated \{112\} plane GBs and cation-Se-terminated \{112\} plane GBs. 
While the former showed Cu depletion and In enrichment, Cu depletion without In 
enrichment was detected for the latter. Yan \emph{et al.}~\cite{yan_PRL_2006}, 
on the other hand, reported no change in the atomic composition near the 
cation-Se-terminated \{112\} plane GBs. This discrepancy could be due to 
different methods employed to study the atomic 
composition~\cite{hetzer_APL_2005, lei_JAP_2007}. Another type of \S3 \ GB, 
namely \S3 (114), has been studied in \CISe\ and \CGSe~\cite{yan_PRL_2007, 
li_ACSNANO_2011, yin_APL_2013, feng_PLA_2014}. Although this type of GB is not 
experimentally identified in \CISe\ and \CGSe, its structure is adopted from GBs in CdTe. 
A common feature of the \S3 \ GBs is that high percentage 
of these GBs in \CISe\ and \CGSe\ are charge neutral~\cite{baier_APL_2011, 
siebentritt_PRL_2006, jiang_APL_2012}. 

Due to the complex nature of thin-film devices, it would be highly desirable to 
achieve new insights from theoretical calculations. Such results can serve as 
common ground between experimentalists and theoreticians leading to a better 
understanding of the properties of devices. Taking advantage of the calculated 
results makes experimentalists able to improve the cells in the laboratory 
scale. Consequently these knowledge can be transferred to the industrial scale. 

The outcomes of the previous first-principles studies on GBs, however, are not 
conclusive ~\cite{persson_PRL_2003, yan_PRL_2006, li_ACSNANO_2011, yan_PRL_2007, 
yin_APL_2013, feng_PLA_2014}. In their pioneering work, Persson and 
Zunger~\cite{persson_PRL_2003}  employed LDA and LDA+U method to study the 
cation-terminated (112) surface of \CISe\ to explain the anomalous 
characteristics of the symmetric GBs. They proposed that the valence band 
maximum (VBM) of the surface is lower than the bulk VBM therefore the GBs act as 
a hole barrier. The Cu vacancy reconstruction at the surface, which lowers the 
surface VBM with respect to the bulk VBM, is in contrast to the results of Yan 
\emph{et al.}~\cite{yan_PRL_2006}. Other authors employed different flavors of 
(local) LDA and (semi-local) GGA to study the \S3 (114) GBs in \CISe\ and \CGSe. 
While some studies~\cite{yan_PRL_2007, li_ACSNANO_2011} suggested that these GBs 
do not create any gap state, the other studies predicted the formation of deep 
gap states in the bandgap of the systems with GBs \cite{yin_APL_2013, 
feng_PLA_2014}. 

The results achieved by (semi-) local methods remain doubtful, even if they are 
in agreement with experimental findings. This is because of the employed methods 
in the aforementioned studies that do not describe the semiconducting nature of 
\CISe\ and \CGSe\ adequately~\cite{hinuma_PRB_2013,pohl_PRB_2013}. Namely, for 
the defect-free \CISe\ and \CGSe\ bulk, the bandgaps predicted by PBE and PBE+U 
are severely underestimated~\footnote{In our calculations, the \CISe\ bulk 
bandgap calculated by PBE+U method is \unit[0.14]{eV}.}. This failure is partly due 
to the shortcomings of (semi-) local methods and partly due to underestimation 
of the anion displacement ($u$) of \CISe~\cite{vidal_PRL_2010}. The size of the 
bandgap is of particular importance when the defect states, which might appear 
in the bandgap, are studied. In light of the shortcomings of the underlying 
methods, the inconsistencies between previous works are not surprising. 
In this work we report on the first-principles hybrid functional study of two 
types of \S3 \ GBs, namely \S3 (112) and \S3 (114), in \CISe\ and \CGSe. 

\section{Methodology}
Hybrid functionals are known as a rather accurate method to study the electronic 
structure of semiconductors. In this approach a portion of the exact exchange 
calculated by Hartree-Fock method is incorporated into the exchange-correlation 
functional calculated by density functional theory (DFT). In the present work, 
we used HSE06 functional~\cite{heyd_JCP_2003} with the fraction of the exact 
exchange set to \unit[30]{\%}~\cite{pohl_PRB_2011,hinuma_PRB_2012,hinuma_PRB_2013}. Using this 
setup, the value of the anion displacement is calculated with an error smaller 
than \unit[0.3]{\%} and the computed bandgaps are \unit[1.0]{eV} and 
\unit[1.6]{eV} for \CISe\ and \CGSe, respectively, in agreement with 
experimental values~\cite{alonso_PRB_2001}. 

All calculations have been performed within the framework of DFT as implemented 
in Vienna Ab-initio Simulation Package (VASP)~\cite{kresse_PRB_1996}. 
We used the projector augmented wave (PAW)~\cite{bloechl_PRB_1994,kresse_PRB_1999} 
method together with a plane-wave cutoff energy of \unit[300]{eV} and a mesh of 
($\mathrm{3 \times 3 \times 1}$) k-points. For density of states (DOS) 
calculations a denser k-mesh has been used. The supercells consist of 128 and 
180 atoms for the \S3 (112) and \S3 (114) GBs, respectively. To construct the 
supercells, the optimized lattice constants and atomic positions of the bulk 
have been used. The distance between periodic supercells separated by vacuum is 
about \unit[30]{\AA}. 

Slab calculations, which are generally employed to represent a material with GBs, 
might suffer from the charge transfer through the slabs due to  dangling bonds at the 
surfaces of the slabs. To quench the dipole moment of the slabs and prevent 
charge transfer, the surface dangling bonds were passivated with hydrogen-like 
pseudoatoms with partial charges. The valency of these hydrogen-like pseudoatoms 
are chosen in such a way that they provide the amount of missing electrons for 
the surface atoms, so they fulfill the octet rule. The position of atoms in the 
outer four atomic layers were fixed to their bulk position to mimic the 
underlying bulk material. Other atoms were fully relaxed until the forces on 
each atom were below \unit[0.01]{eV/\AA}. Using the same methodology, a 
defect-free supercell with \{112\}-plane termination was constructed and 
considered as the reference. 

\begin{figure}[!t]
  \includegraphics[width=0.40\textwidth]{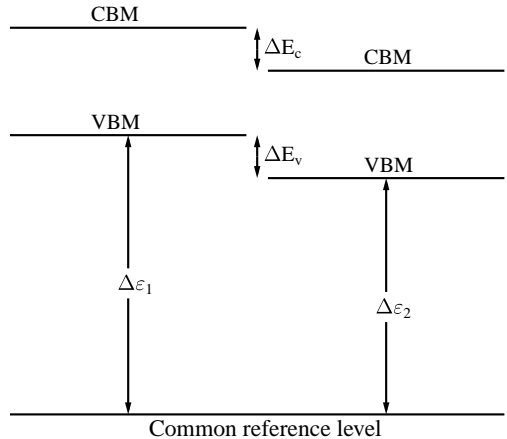}
  \caption{Band offset between system \textquoteleft1\textquoteright  and \textquoteleft2\textquoteright.
  The common reference level is the average electrostatic potential of the periodic bulk.}
    \label{fig:bandalignment}
\end{figure}

To evaluate the relative shift of the VBM and conduction band minimum (CBM) 
with respect to the bulk, the macroscopic average of the electrostatic potentials are 
calculated~\cite{dandrea_JVST_1993, 
alkauskas_pss_2011,damico_APL_2012,komsa_PRB_2012}. The VBM shift can be 
expressed as 
\begin{equation} \Delta E_{\sss v} = \Delta \varepsilon_{\sss 1} - 
\Delta \varepsilon_{\sss 2}, 
\end{equation} 
where $\Delta \varepsilon_{\sss 1}$ is the energy difference from the VBM to the 
reference level of system \textquoteleft1\textquoteright~ and $\Delta 
\varepsilon_{\sss 2}$ is the energy difference from the VBM to the reference 
level of system \textquoteleft2\textquoteright, see 
FIG.~\ref{fig:bandalignment}. The energy difference  $\Delta \varepsilon_{\sss 
1}$ and $\Delta \varepsilon_{\sss 2}$ are calculated in the region far form the 
GBs and are very close the values calculate for the perfect periodic bulk. By 
knowing the size of the bandgap, the CBM shift calculation is straightforward.

The crystal structure of \CIGSe\ and CdTe has the same fundamental 
characteristics and the GBs in \CISe\ and \CGSe\ can be modeled based on the 
observed GB structures in CdTe. In the present work, \S3 (112) and \S3 (114) 
correspond to lamellar and double-positioning twins in CdTe~\cite{yan_JAP_2001, 
yan_JAP_2003}. In the \S3 (112) GBs, either a cation-containing plane is next to 
a Se-containing plane (\S3 (112)-I) or Se-containing planes are facing each 
other (\S3 (112)-II). In the \S3 (114) GBs, either Se atoms have dangling bonds 
(\S3 (114)-I) or cations have dangling bonds (\S3 (114)-II). The  atomic 
structure of the \S3 (112) and \S3 (114) GBs are very different: while the \S3 
(114) GBs contain dangling, wrong, and extra bonds the atomic structure of \S3 
(112)-I is very similar to the bulk and the \S3 (112)-II GB contains Se dangling 
and Se-Se wrong bonds.

\begin{figure}[!b]
\includegraphics[width=0.40\textwidth]{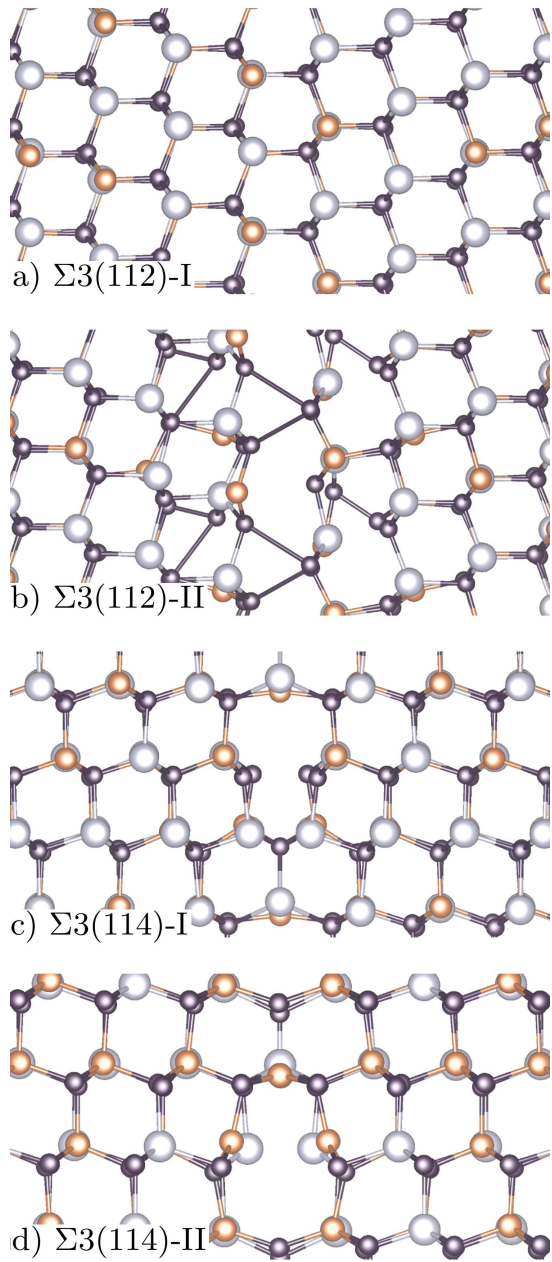}
\caption{(Color online) Atomic structure of the (a) cation-Se-terminated (type I) and 
 (b) Se-Se-terminated (type II) \S3 (112) GB. 
(c) and (d) depict the atomic structure of the type I and type II of the \S3 (114) GB. 
The Cu, In, and Se atoms are shown as brown, large-light gray, and small-dark gray spheres, respectively.}
\label{fig:GBs}
\end{figure} 

\section{Results and discussion}
The atomic structures of GBs in \CISe\ after geometry optimization are presented 
in FIG.~\ref{fig:GBs}. While \S3 (112)-I shows little changes in its structure, 
\S3 (112)-II undergoes dramatic structural relaxation. Compared to the bulk, Se 
atoms at the \S3 (112)-II GB are not surrounded by four cations. After 
optimization, the minimum distance between Se atoms increases from \unit[3.11]{\AA} 
to \unit[3.52]{\AA}. In contrast to \CISe, Se atoms at the \S3 (112)-II GB of \CGSe\ 
(not shown here) are not located at the outermost layer of the GB: on 
one side of the GB Se atoms migrate into the bulk and instead of Se-Se 
wrong bonds at the GBs, Se-cation bonds are formed. In a qualitative agreement 
with the previous studies~\cite{yan_PRL_2007, yin_APL_2013}, in the \S3 (114) 
GBs the atoms with dangling bonds show larger relaxation compared to the other atoms. 

\begin{figure}[!b]
\includegraphics[width=0.40\textwidth]{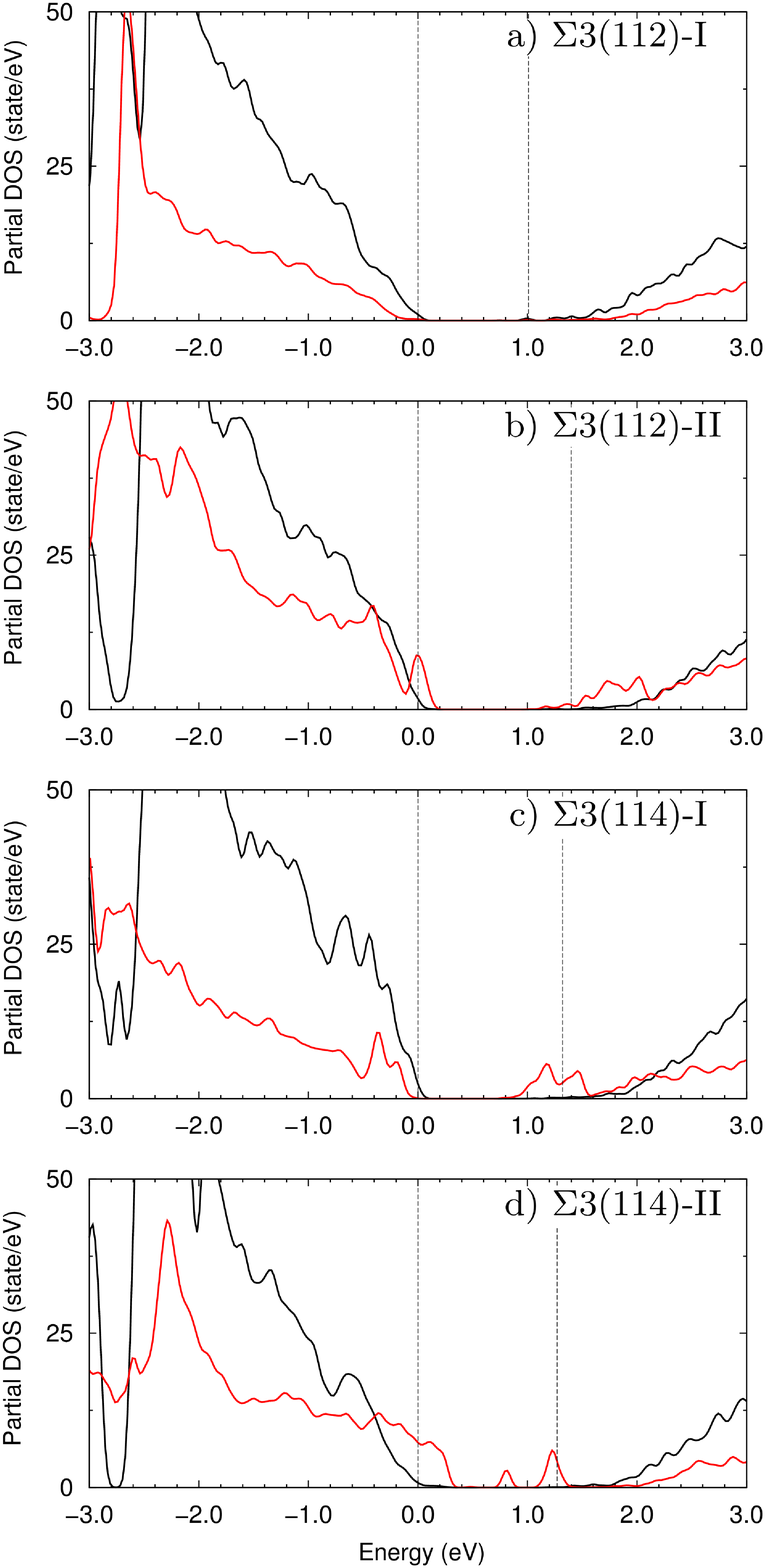}
\caption{(Color online) Partial DOS calculated for GBs in \CISe.
Black and red line show the projected DOS for the bulk atoms and the atoms close to the GBs, respectively.
The zero of the energy is set at the bulk VBM. Vertical dashed lines show the VBM and CBM.}
\label{fig:DOSCIS} 
\end{figure}

It has been discussed that dangling and wrong bonds in CdTe can create defect 
levels~\cite{zhang_PRL_2008}. Since the \S3 (114) and \S3 (112)-II GBs contain 
these defects, one expects to see defect levels in the bandgap of the systems 
with these GBs as well. To study the effects of the GBs on the electronic 
structure, we calculated atom-projected partial DOS for the systems with GBs. 
The projected partial DOS for the \S3 (112) and \S3 (114) GBs in \CISe\ are 
shown in FIG.~\ref{fig:DOSCIS}. All GBs (except \S3 (112)-I that creates no gap 
state) create unoccupied gap states close to the CBM which are in resonance with 
the conduction band states in agreement with the $dI/dU$ 
simulations~\cite{moenig_PRL_2010}. From the total energy of the charged ($+1$ 
and $-1$) GBs we have computed the thermodynamic charge transition levels. In 
\CISe\ and \CGSe, the positively-charged GBs are not stable but the defect 
states can trap electrons to make the GBs negatively charged.

\begin{figure}[!b]
\includegraphics[width=0.40\textwidth]{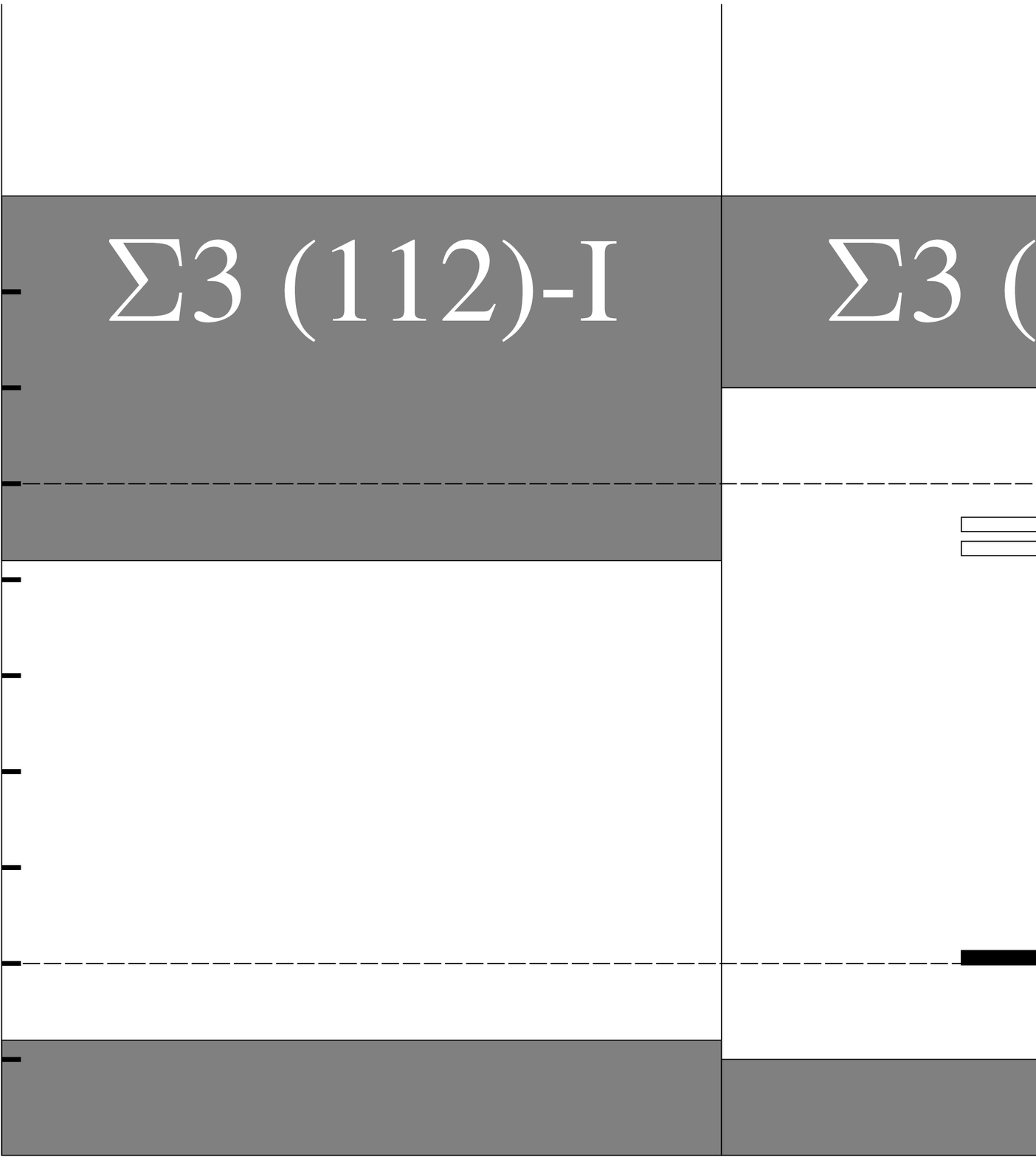}\\[20pt]
\includegraphics[width=0.40\textwidth]{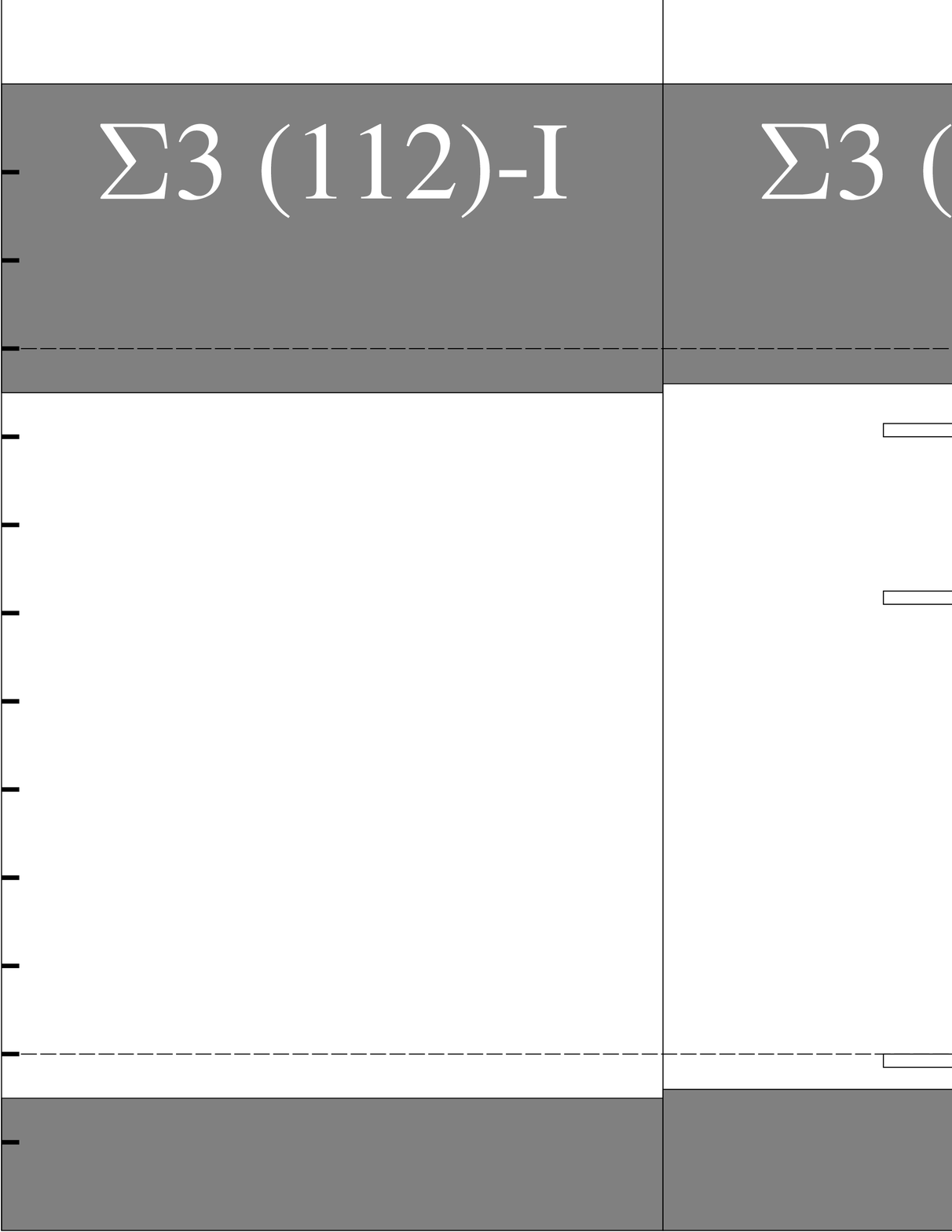}
\caption{Schematic band offset and gap 
states of GBs for \CISe\ (top panel) and \CGSe\ (bottom panel).
The occupied and unoccupied single particle levels are represented as black and white
rectangles. Horizontal dashed lines show the bandgap of the bulk \CISe\ 
($E_{\sss g} = \unit[1.0]{eV}$) and \CGSe\ ($E_{\sss g} = \unit[1.6]{eV}$), respectively.
The distance between tics is \unit[0.2]{eV}}
\label{fig:schematic_CISe_CGSe}
\end{figure}

Our data show that the existence of GBs shifts the VBM and CBM with respect to 
the bulk VBM and CBM. This can lead to `electron-free'/`hole-free' zone near the 
GBs where the concentration of electrons/holes is less than in the grain 
interior (GI). It has been discussed that the creation of such a barrier at the 
GBs for one type of the carrier (electron or hole) impedes electron-hole 
recombination at the GBs~\cite{persson_PRL_2003, persson_APL_2005, gloeckler_JAP_2005}. 
FIG.~\ref{fig:schematic_CISe_CGSe} schematically presents the computed band 
offsets between the GB and GI for \CISe\ and \CGSe. 

The other effect of the GBs on the electronic structure is to change the size of 
the bandgap. In the case of \CISe, the systems containing \S3 (112)-I, \S3 
(112)-II, \S3 (114)-I, and \S3 (114)-II have the bandgap size of \unit[1.0]{eV}, 
\unit[1.4]{eV}, \unit[1.3]{eV}, and \unit[1.3]{eV}, respectively. For the 
corresponding GBs in \CGSe\ the bandgaps are \unit[1.6]{eV}, \unit[1.6]{eV}, 
\unit[1.8]{eV}, and \unit[1.8]{eV}, respectively. In the following paragraphs, 
we give a detailed discussion of our findings for different types of GBs. 

$\mathbf{\Sigma 3} \textrm{\textbf{(112)-I:}}$
in \CISe\ and \CGSe, the \S3 (112)-I GBs do not create gap states but shift the 
VBM and CBM with respect to the bulk (FIG.~\ref{fig:schematic_CISe_CGSe}). This 
band alignment draws electrons to the GB region but reflects holes away. The 
electron-hole recombination remains low because of insufficient holes at the GB. 
This GB forms easily in \CIGSe\ due to its highly symmetric 
structure~\cite{baier_PhD_2012} and is harmless for the carrier transport.

$\mathbf{\Sigma 3} \textrm{\textbf{(112)-II:}}$
the \S3 (112)-II GB in \CISe\ creates three gap states: one occupied state which 
is \unit[0.1]{eV} above the VBM and two unoccupied states which are 
\unit[0.3]{eV} and \unit[0.4]{eV} below the CBM. The $\varepsilon(+/0)$  level 
is not in the bandgap, therefore holes are not trapped in the defect states. The 
$\varepsilon(0/-)$ level is in the bandgap, positioned \unit[0.2]{eV} below the 
CBM. This defect level can be occupied by electrons only if \CISe\ is 
\mbox{n-type}. Our results are in agreement with the experimentally-observed 
charge-neutral \S3 (112) GBs in \mbox{p-type} \CISe~\cite{baier_APL_2011}. In 
the system containing this type of GB, the VBM and CBM are \unit[0.2]{eV} 
lower and higher than the bulk VBM and CBM, respectively (see 
FIG.~\ref{fig:schematic_CISe_CGSe}, top panel). This band alignment expels 
electrons and holes from the GB region. These electronic properties of the \S3 
(112)-II GBs in \CISe\ explain why this type of GB is harmless for the carrier 
transport in \mbox{p-type} \CISe. 

In the case of \CGSe, the $\varepsilon(+/0)$ level is not in the bandgap but 
$\varepsilon(0/-)$ is \unit[1.5]{eV} below the CBM. Hence, electrons from 
the conduction band can be trapped in this level. Considering the band alignment 
(FIG.~\ref{fig:schematic_CISe_CGSe}, bottom panel), both the VBM and CBM of this 
system are \unit[0.2]{eV} lower than the VBM and CBM of the bulk. That is, while 
this GB creates an electrostatic barrier for holes, concentration of electrons 
close to this GB is higher than in the GI. The valence band offset, however,  is 
not large enough to suppress the holes concentration at the 
GB~\cite{gloeckler_JAP_2005} therefore the carrier lifetime is reduced if this 
type of the GB is formed in \CGSe.

$\mathbf{\Sigma 3} \textrm{\textbf{(114)-I:}}$ 
as it is shown in FIG.~\ref{fig:DOSCIS} (c), the dangling and wrong bonds of the 
\S3 (114)-I GB in \CISe\ cause unoccupied gap states to form. The deepest 
unoccupied state is \unit[0.5]{eV} below the CBM. The $\varepsilon(0/-)$ and 
$\varepsilon(+/0)$ charge transition levels are \unit[0.7]{eV} and \unit[0.5]{eV} 
below the CBM and VBM, respectively. Thus, the positively-charged gap states 
are not stable but the defect levels can trap electrons and become negatively 
charged. The band offset of this system, however, screens this GB from bulk 
carriers and results in low probability of recombination.

Formation of the \S3 (114)-I GBs in \CGSe\ creates three unoccupied gap states. 
The $\varepsilon(0/-)$ level is \unit[1.2]{eV} below the CBM, therefore the 
defect levels can trap electrons even for a \mbox{p-type} \CGSe. Holes, on the 
other hand, cannot be trapped in the defect levels. The conduction band offset 
(FIG.~\ref{fig:schematic_CISe_CGSe}, bottom panel) makes the region close to 
this GB electron rich. The motion of holes into the GB region, on the other 
hand, is impeded due to the valence band offset. The size of the valence band 
offset, however, is not large enough to mitigate the effect of the presence of 
this GB in the system~\cite{gloeckler_JAP_2005}.

$\mathbf{\Sigma 3} \textrm{\textbf{(114)-II:}}$
the cation dangling bonds create both occupied and unoccupied states in the bandgap of 
\CISe\ and \CGSe. For \CISe\ (see FIG.~\ref{fig:DOSCIS} (d)), the occupied gap 
states are close to the VBM and the deepest state is \unit[0.2]{eV} above the 
VBM. The deepest unoccupied gap state is \unit[0.1]{eV} below the CBM. The 
$\varepsilon(+/0)$ level is not in the gap and $\varepsilon(0/-)$ is 
\unit[0.5]{eV} below the CBM. This means, positively-charged gap states 
are not stable but if the chemical potential of the electrons is high, i.\,e.\@ 
\mbox{n-type} \CISe, then electrons can be trapped in the defect levels. Still, 
the electron- and hole-free zone near this GB make the probability of recombination low.

The VBM and CBM of this GB in \CGSe\ are \unit[0.3]{eV} and \unit[0.1]{eV} below 
the bulk VBM and CBM, respectively. The transition level $\varepsilon(+/0)$ is 
not in the gap and $\varepsilon(0/-)$ is \unit[0.9]{eV} below the CBM. This GB 
remains neutral for \mbox{p-type} \CGSe\ but for high chemical potential of the 
electrons, the gap states can become negatively charged. The conduction band 
offset for this GB (FIG.~\ref{fig:schematic_CISe_CGSe}, bottom panel) also leads 
to a high concentration of electrons close to the GB, meaning that the chance of 
electrons to be trapped in the gap state is high. Although the valence band 
offset repels holes from the GB, it cannot effectively screen the GB from holes 
and this GB is prone to recombination~\cite{gloeckler_JAP_2005}.

We note that in our study the electronic-structure changes are merely due to the 
existence of the GBs. To look into the influence of changing the chemical 
potential of the constituent atoms, we have studied the effect of the formation 
of a charge-neutral defect pair ($2\mathrm{V}_{\sss Cu}^{-}+ \mathrm{In}_{\sss 
Cu}^{++}$) on the electronic structure of {\S3 (112)-II} in \CISe. In agreement 
with previous results~\cite{persson_PRL_2003}, the VBM is lower than the bulk 
VBM. In this system, the occupied gap state is removed from the bandgap but the 
unoccupied gap states are still present.

\section{Summary}
In summary, our results are the following: 
(i) to study GBs in \CIGSe, employing a method that can 
correctly describe the electronic structure of this material, hybrid functional for example, is essential. 
(ii) The formation of GBs can alter the electronic structure of their systems in three different ways: 
a) GB creates gap states,
b) GB shifts the VBM and CBM with respect to the bulk, and 
c) GB changes the bandgap size.
(iii) Although gap states can be detrimental for the carrier transport, 
the band offsets and the change in the bandgap sizes mitigate this destructive effect in \CISe.
(iv) The behavior of the \S3 \ GBs in \CGSe \ is different from \CISe. The conduction band offset draws electrons 
to the GB region and the valence band offset is not large enough to suppress the concentration of holes. This band offset
makes the presence of the symmetric GBs in \CGSe\ destructive, except for the \S3 (112)-I GB. 

The electrically benign behavior of the \S3 \ GBs in \CISe\ suggests that the \S3 \ 
GBs in this material do not need passivation. The detrimental behavior GBs in 
\CGSe, on the other hand, shows the necessity of the GBs passivation.
Our findings could be of interest to the research groups seeking to optimize the 
device fabrication and to improve the efficiency of the solar cells based on \CIGSe.

\section{Acknowledgment}
We acknowledge the financial support from the 
\emph{Bundesministerium f\"ur Wirtschaft und Energie (BMWi)} 
for the comCIGS~II project (0325448C).

\end{document}